\documentclass[12pt,preprint]{aastex}

\slugcomment{To be submitted to ApJ Letter}

\newcommand{\lapprox}{{\footnotesize $\buildrel < \over \sim \,$}}
\newcommand{\gapprox}{{\footnotesize $\buildrel > \over \sim \,$}}

\shorttitle{Crystalline Silicates in IRAS\,17495+2534}
\shortauthors{Speck et al.}

\begin{document}
\title{The Cosmic Crystallinity Conundrum: Clues from IRAS\,17495$-$2534}

\author{Angela K.\ Speck\altaffilmark{1,2},
Alan G. Whittington\altaffilmark{2}
Josh B. Tartar\altaffilmark{1}
}

\altaffiltext{1}{Department of Physics and Astronomy, University of Missouri, 
Columbia, MO 65211}
\altaffiltext{2}{Department of Geological Sciences, University of Missouri, 
Columbia, MO 65211}

\begin{abstract}
Since their discovery, cosmic crystalline silicates 
have presented several challenges to understanding dust formation and 
evolution.
The mid-infrared spectrum of IRAS\,17495$-$2534, a highly obscured 
oxygen-rich asymptotic giant branch (AGB) star, is the only source observed to 
date which exhibits a clear 
crystalline silicate absorption feature. 
This provides an unprecedented opportunity to test 
competing hypotheses for dust formation. 
%
Observed spectral features suggest that both amorphous and crystalline 
dust is dominated by forsterite (Mg$_2$SiO$_4$)
rather than enstatite (MgSiO$_3$) or other silicate compositions.
We confirm that high mass-loss rates should produce more crystalline material, 
and show why this should be dominated by forsterite. The presence of
Mg$_2$SiO$_4$ glass suggests that another factor (possibly C/O) is 
critical in determining astromineralogy.
Correlation between crystallinity, mass-loss rate and initial 
stellar mass suggests that only the most massive AGB stars contribute 
significant quantities of crystalline material to the interstellar medium, 
resolving the conundrum of its low crystallinity. 
\end{abstract}





\section{Introduction}
\label{intro}


One of the most exciting recent developments in astronomy 
was the discovery of crystalline silicate stardust by the Infrared Space 
Observatory \citep[ISO;][]{isoref}.
Crystalline silicates were initially discovered around 
evolved intermediate mass stars at far-infrared (IR) wavelengths 
\citep{waters96}, but
have since been detected in 
young stellar objects \citep{herbig}, comets \citep{comets}, and 
Ultra Luminous Infrared Galaxies \citep{ulirg}. 
In the interstellar medium (ISM) silicate grains in crystalline form are rare 
\citep[$\sim$1\% crystalline by mass;][]{min07,kemper04}. 
However, Asymptotic Giant Branch (AGB) stars and 
their successors (pre-planetary and planetary nebulae) are the major 
contributors of material to the ISM \citep{kwok04}. It has been suggested 
that AGB stars could contain \gapprox 10\% crystalline silicates 
\citep{kemper04}, 
so it is vital to understand the formation 
of dust around AGB stars
to resolve this apparent discrepancy.
%


AGB stars are luminous cool giants, with high mass-loss rates 
($10^{-7} < \dot{M} < 10^{-3}$\,M$_{\odot}$/yr), which increase over time 
\citep{mloss}. They are highly evolved descendants of low- 
to intermediate-mass (0.8--8\,M$_{\odot}$) stars. 
%
Mass loss produces a circumstellar envelope, where dust forms. 
The order in which different species condense from the outflowing gas 
depends on the physical conditions 
within the envelope. 
Once $\dot{M}$ is very high the circumstellar shell 
becomes optically thick, even at IR wavelengths, and the silicate 
emission features absorb themselves.
The dust mineralogy is dictated by the dust condensation sequence and depends 
strongly on $\dot{M}$ \citep{dijk05}. 
\citet{blommaert07} showed that the mineralogy depends only on the age of the 
star for a given initial stellar mass.

To date, crystalline silicate features have been seen clearly only in emission.
It has been proposed that crystalline silicate absorption at 11$\mu$m
contributes to the broadening of the classic 10$\mu$m amorphous silicate 
absorption feature associated with OH/IR stars \citep{sylvester98,vanhollebeke}.
However, IRAS\,17495$-$2534 (hereafter I17495) 
is the first object to show a distinct crystalline 11$\mu$m 
absorption feature.

\begin{figure}[t]
\includegraphics[angle=270,scale=0.7]{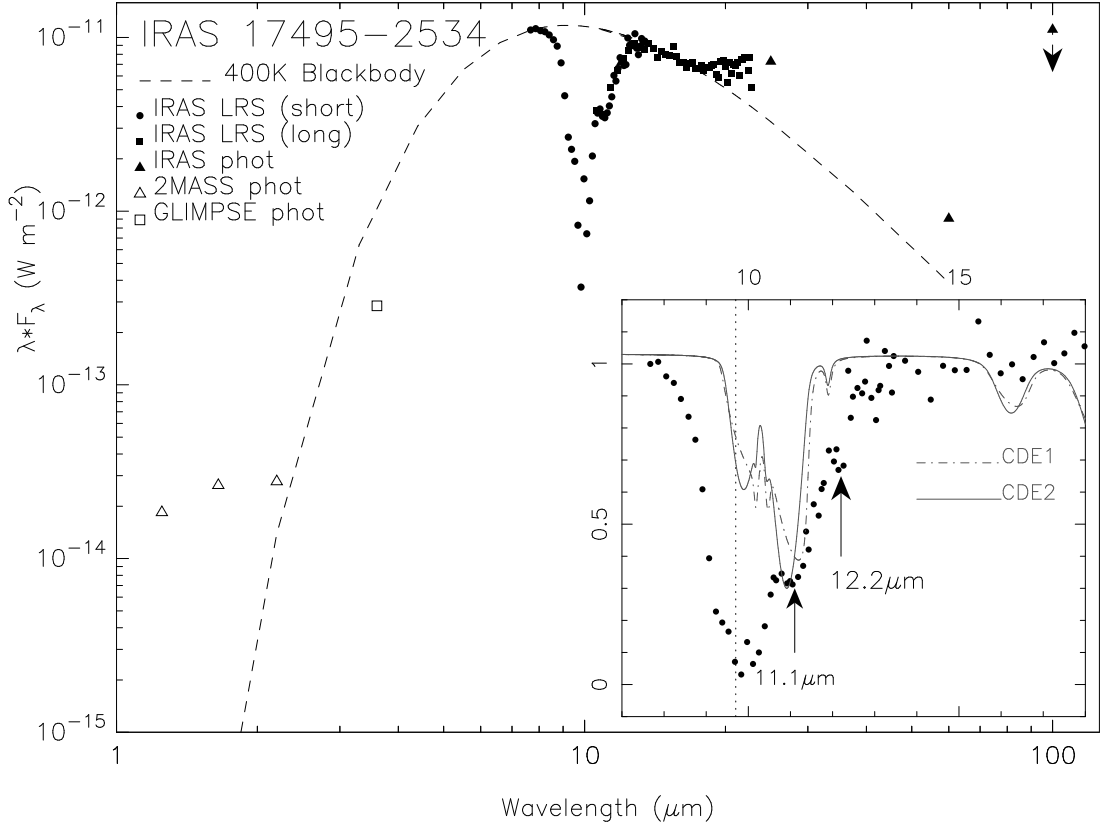}
\caption{\label{SED} The spectrum of IRAS\,17495$-$2534:
$y$-axis is flux ($\lambda F_\lambda$) in W\,m$^{-2}$.
{\it Main panel}: 
log-log plot of the full spectral energy distribution (SED) including
the IRAS LRS spectrum, 
the near-IR JHK photometry from 2MASS, 
3.6$\mu$m photometry form the Spitzer GLIMPSE survey, and 
far-IR photometry points from IRAS. 
{\it Inset}:
linear-scale close-up on the continuum-divided IRAS LRS spectrum, 
with crystalline features indicated by arrows. 
$y$-axis is normalized flux.
Grey lines are calculated mass absorption coefficients for continuous 
distributions of ellipsoidal (CDE) crystalline forsterite grains 
from \citet{fabian}.
CDE1 has equal probability of all shapes;
CDE2 has quadratic weighting, such that near-spherical particles
are most probable.
Dotted line at 9.7$\mu$m is
included to demonstrate the redness of the peak of the absorption feature.}
\end{figure}

I17495 is located in the Galactic Plane (gal.\ coord.\ = 003.6844$+$00.3880) 
at a distance of $\sim$4kpc \citep{loup93}, about half way to the Galactic 
Center. The IRAS LRS spectrum clearly exhibits an 11.1$\mu$m absorption 
feature superposed on the classic 10$\mu$m amorphous silicate feature
(Fig.~\ref{SED}).
This is the first clear crystalline silicate absorption feature 
seen to date in any object. 

$\dot{M}$ for this object is estimated to be $\sim 2 \times 10^{-4} < \dot{M} 
< 5\times 10^{-4}$\,M$_{\odot}$/yr based on  mid-IR ([25]-[12]) color 
\citep{loup93} and the 
range of optical depths ($15 < \tau_{10 \rm \mu m} < 45$) consistent 
with our model results (see \S~\ref{RTmod}). 
Both $\dot{M}$ and
expansion velocity \citep[$v_{\rm exp}$, 16km/s;][]{loup93} 
are at the high end of normal for AGB stars. 
Most extreme O-rich AGB stars are OH/IR stars, exhibiting OH-maser emission.
I17495 is one of a handful of visibly obscured O-rich AGB stars 
not exhibiting OH-maser emission \citep[dubbed color-mimics;][]{bmlewis1}. 
However, the spectra of other color mimics more closely resemble 
those of OH/IR stars, i.e., when observed, the putative 11$\mu$m crystalline 
silicate feature is merely a hump superposed on the regular amorphous silicate 
absorption feature.
This is also true for OH/IR stars with similar mass-loss rates in the 
Galactic Bulge (GB). 
Even the GB source with the highest $\dot{M}$
\citep[$3\times 10^{-4}$M$_\odot$/yr; IRAS\,17276$-$2846;][]{vanhollebeke} 
shows the 11$\mu$m feature as a hump, rather than the distinct 11$\mu$m 
feature seen in Fig.~\ref{SED}.
%

While forsterite (crystalline Mg$_2$SiO$_4$) is expected to have a peak near 
11.3$\mu$m 
\citep[e.g.][]{koike}, the peak can shift to shorter wavelengths. 
In the laboratory, \citet{jaeger2} found that synthetic 
forsterite peaks closer to 11.2$\mu$m, and matches the position of the hump 
observed in GB OH/IR stars \citep{vanhollebeke}. 
\citet{fabian} investigated grain shape effects. Their mass absorption 
coefficients for ellipsoidal forsterite grains are included in Fig.~\ref{SED}, 
and demonstrate that the crystalline forsterite feature can peak at 
11.1$\mu$m. 
\citet{boersma} attributed an 11.1$\mu$m emission feature in the 
spectrum of a Herbig Ae/Be star to forsterite. 
\citet{tamanai} found that, for non-embedded free-flying particles of 
forsterite, the feature actually peaks at 11.06$\mu$m.
Composition, grain-shape and grain-size can significantly shift the 
position of this feature.
%

\section{Crystallinity \& Mass-loss rates}

Crystalline silicates were first observed around evolved stars with very 
high $\dot{M}$, leading to the inference that crystal formation 
requires such conditions \citep[e.g.][]{cami98}.
However, the diagnostic infrared features of crystalline grains can remain 
hidden by the strong amorphous silicate features until 
$\dot{M} \approx 10^{-5}$\,M$_{\odot}$/yr \citep{kemper01}. 

\begin{figure}[t]
\includegraphics[angle=270,scale=0.7]{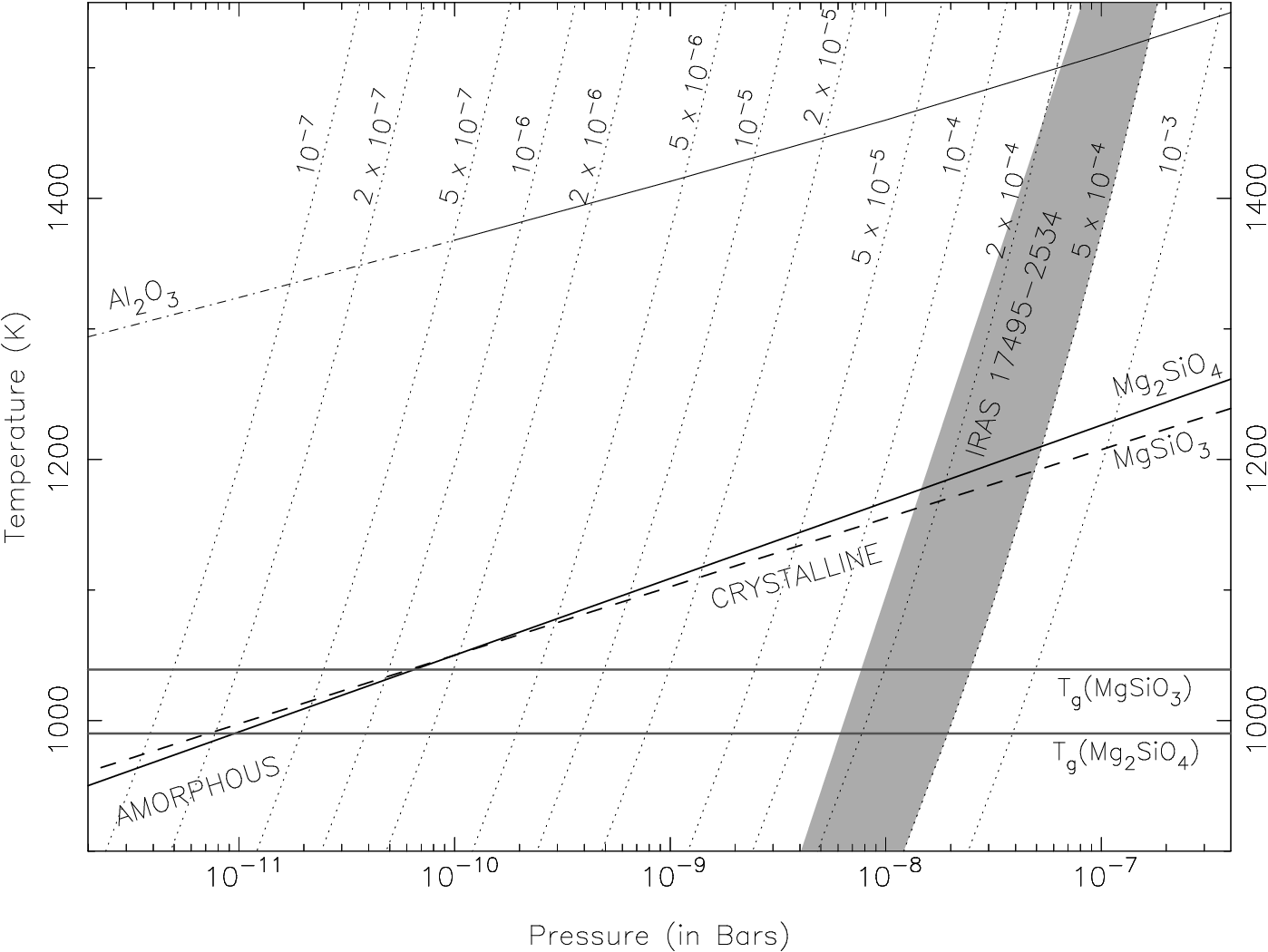}
%
\caption{\label{PTspace} P--T space for dust condensation around O-rich AGB 
stars:
Solid and dashed lines indicate $T_{\rm dust}$ for a given pressure from 
thermodynamic equilibrium calculations (relevant compositions are labeled) 
\citep{lf99}.
The stability temperature of Al$_2$O$_3$ is dot-dashed where extrapolated. 
For all $\dot{M}$ values, Al$_2$O$_3$ forms at a significantly higher 
temperature 
than the silicates, and thus can form a seed nucleus.
Dotted lines indicate the P--T paths for the outflowing gas for a range of 
$\dot{M}$ (indicated in M$_\odot$/yr) 
as calculated from $\dot{M}$ and $v_{\rm exp}$. 
Thick grey horizontal lines indicate $T_g$ for Mg$_2$SiO$_4$ and MgSiO$_3$.
Grey shaded area shows the P--T space relevant to I17495.}
\end{figure}

Dust condensation temperature ($T_{\rm dust}$) are depressed at 
low gas pressures \citep{lf99,gs99}.
Fig.~\ref{PTspace} shows the calculated pressure-temperature (P--T) conditions 
in the 
dust-forming zone around O-rich AGB stars \citep[see][for method]{dijk05}, 
compared to thermodynamic equilibrium dust condensation models \citep{lf99}, 
and the glass-transition temperatures ($T_g$) for Mg$_2$SiO$_4$ 
\citep{richet} and MgSiO$_3$ \citep{wilding}. 
Grains forming below $T_g$ will be amorphous, 
while grains forming much above this temperature 
should crystallize rapidly \citep{ks99}.
For $\dot{M}$ \lapprox $\sim 2 \times 10^{-7}$ M$_{\odot}$/yr, $T_{\rm dust}$
for both Mg$_2$SiO$_4$ and  MgSiO$_3$ fall below their respective $T_g$, thus 
precluding crystal formation.
Above this threshold $\dot{M}$, dust grains should form at temperatures 
where crystalline structures would form either directly, or by annealing
of initially amorphous grains.
Once dust forms, the radial temperature profile steepens, because of 
absorption of starlight by dust grains. Including this 
factor, we can use the outflow velocity to calculate the timescale for dust to 
cool from $T_{\rm dust}$ to  $T_g$. For $\dot{M} > 10^{-5}$\,M$_{\odot}$/yr, 
the timescale is $\sim$6--12 months, providing ample time for annealing 
and crystallization.
This suggests that crystalline silicates should be more common in high 
$\dot{M}$ stars and the observed $\dot{M}$--crystallinity correlation is not 
merely due to masking of the crystalline spectral features 
\citep[c.f.][]{ks99}.
In addition, for stars with $\dot{M}$\gapprox $10^{-5}$\,M$_{\odot}$/yr 
forsterite forms at a higher temperature than enstatite 
(crystalline MgSiO$_3$), and the difference in $T_{\rm dust}$ for these two 
species increases with $\dot{M}$ (see \S~\ref{dustformmech}).

\section{Radiative transfer (RT) modeling}
\label{RTmod}

We used the 1-D radiative transfer (RT) program DUSTY
\citep{ie95}, to determine the gross mineralogy associated with the mid-IR 
absorption features.
RT modeling is inherently degenerate \citep{ie97}
and hence models are not unique. 
Consequently, we use the modeling only to constrain $ \tau_{10 \rm \mu m}$ and 
to determine which minerals are inconsistent with the observed spectral 
features. 
A detailed analysis of the model results will be presented elsewhere.

At pressures relevant to I17495, 
$T_{\rm dust}\approx$1200\,K for  Mg$_2$SiO$_4$ and  MgSiO$_3$.
Models with T$_{\rm dust} <$ 800\,K could not match the observations 
because the near-IR becomes too bright, and in the 
mid-IR the spectrum slopes the wrong way.
One important result of the modeling is that pyroxenes can be excluded from 
the mineralogy in I17495. The underlying amorphous silicate feature is 
unusually narrow and peaks closer to 10$\mu$m rather than 9.7$\mu$m, 
consistent with Mg$_2$SiO$_4$ glass 
\citep[][and references therein]{ohm92}.
Models incorporating amorphous silicate of pyroxene composition
\citep[e.g.][]{jaeger} or other more silica-rich compositions 
\citep[e.g.][]{ohm92} are inconsistent with the observed spectrum,  
as the peak position of the feature gets too blue, and the FWHM gets too wide. 
Increasing the iron content in olivine has a similar effect on the peak 
position \citep{ohm92}.
Our models suggest that the crystalline component is also dominated by 
forsterite with enstatite and other 
crystalline compositions limited to less than a few percent of the dust mass. 
Therefore, both amorphous and crystalline dust is dominated by Mg$_2$SiO$4$.
This observed forsteritic mineralogy needs to be understood in terms of 
dust formation mechanisms.

\section{Competing dust formation mechanisms}
\label{dustformmech}

There are effectively three competing dust formation mechanisms for 
circumstellar environments:
(i) thermodynamic equilibrium condensation \citep{lf99}; 
(ii) formation of chaotic solids in a supersaturated gas followed by 
annealing \citep{nuth};
(iii) formation of seed nuclei in a supersaturated gas, followed by mantle 
growth \citep{gs99}. 
The latter should follow thermodynamic equilibrium as long as density is high 
enough for gas-grain reactions to occur.

Several observational studies support the thermodynamic condensation sequence 
\citep{dijk05,blommaert07}, which is consistent with both (i) and (iii).
In mechanism (ii), chaotic grains form with the bulk composition of the gas, 
and then anneal if the temperature is high enough \citep{nuth}. 
This mechanism predicts that at low C/O ratios, the dust grains would comprise 
a mixture of olivine, pyroxene and silica, rather than be dominated by olivine 
alone. At high C/O ratios, Al-O bonds are predicted to form preferentially, 
leading to dust dominated by oxides rather than silicates. These predictions 
are inconsistent with observations \citep{dijk05,blommaert07}.

In the seed-mantle model, Mg$_2$SiO$_4$ is expected to use more refractory 
condensates as nucleation centers (e.g.\ Al$_2$O$_3$; 
see Fig.~\ref{PTspace}). 
Whether the transformation into MgSiO$_3$ by gas-grain reaction
occurs is then dependent on gas density. Condensation and annealing 
experiments show that the resulting dust is dominated by forsterite;
enstatite is present only in trace amounts \citep[e.g.][]{dd78,nd82,nuth2}. 
In laboratory experiments, the density is always high compared to 
circumstellar shells, so they may only be relevant to the highest $\dot{M}$.

The modeled mineralogy is consistent with an abrupt decrease in density 
between the formation temperatures for forsterite and enstatite. 
Once $\sim$20\% of the Si atoms are incorporated into dust, the opacity is 
such that radiation pressure accelerates the circumstellar material,
resulting in a precipitous density decrease, and inhibiting further 
gas-grain reactions \citep{gs99}. 
At high $\dot{M}$, forsterite is stable at a higher 
temperature than enstatite. Formation of forsterite grains provides the 
impetus for radiation-driven acceleration and thus reduces the efficacy of 
enstatite formation, leaving gas that is depleted in Mg and thus 
more silica-rich. 
Without a surface on which to nucleate, silicates will tend to form amorphous 
solids rather than crystals \citep{microg}, thus the seed nuclei are of the 
utmost importance to forming crystalline grains. Moreover, the dispersion due 
to radiation pressure may lead to seedless amorphous dust formation in the low 
density gas, which should be slightly more silica-rich than MgSiO$_3$ due to 
the depletion of Mg into forsterite grains. The glass composition 
should tend towards Mg$_2$Si$_3$O$_8$, which is one of the metastable 
compositions predicted by \citet{rietmeijer}.

It is possible that the appearance of the clear crystalline absorption feature 
in I17495 is related not to increased crystallinity, but rather to less 
masking of the feature as a result of similar crystalline and amorphous 
silicate compositions \citep[see][]{kemper01}. 
This still leaves the question of why this source is different --
how can this source have both its crystalline and amorphous dust components 
dominated by Mg$_2$SiO$_4$, given that the bulk composition of the gas is 
closer to MgSiO$_3$?

\section{Factors influencing crystallinity and mineralogy}
\label{factors}

%
The physical factors which could give rise to the 
extremely high crystal content include:
(1) $\dot{M}$; 
(2) $v_{\rm exp}$; 
(3) metallicity; 
(4) C/O ratio. 
$\dot{M}$ for I17495 is high, but similar to OH/IR stars and other color 
mimics, suggesting that $\dot{M}$ alone cannot explain the high degree of 
crystallinity exhibited by this source.
A low $v_{\rm exp}$ would increase the pressure at a given temperature 
and thus promote grain formation/growth at higher temperatures. However,  
$v_{\rm exp}$ for I17495 is relatively high.
Increasing metallicity increases the partial pressure of the dust forming 
elements, and thus increases the dust formation temperature at a given total 
gas pressure. The metallicity of the source is unknown, although its location 
in the Galactic plane suggests that it may have higher than solar metallicity. 
If metallicity were the determining 
cause, we would expect to see similar spectra in the GB OH/IR stars. The fact 
that even the high $\dot{M}$ GB stars do not exhibit such a distinct 
crystalline silicate feature suggests that a combination of $\dot{M}$ and 
metallicity would also fail to increase the crystal fraction.

Having eliminated $\dot{M}$, $v_{\rm exp}$ and metallicity as likely causes 
of higher crystallinity, we are left with C/O ratio as the most likely culprit.
The condensation scheme suggested in \S~\ref{dustformmech} is relevant to 
solar C/O ($\approx$0.48). Oxygen-rich AGB stars have C/O ratios between 0.4 
and 0.8, where the exact value depends on both the initial mass and age of the 
star; increasing this ratio decreases the availability of oxygen atoms with 
which to make dust.

The Mg/Si ratio in the circumstellar outflow from which the dust forms is 
close to unity. The silica-poor composition of the amorphous dust 
requires sequestration of some silicon into a non-silicate phase to increase 
the Mg/Si ratio \citep[c.f.][]{fg01}. 
While the formation of gehlenite (Ca$_2$Al$_2$SiO$_7$) before the Mg-rich 
silicates is predicted in many condensation models \citep[e.g][and references 
therein]{lf99,gs99,tielens90}, Si/Ca $\sim$16, and thus gehlenite 
is not an effective sink for silicon atoms.

Other options would be keeping the Si in gas phase SiO; or condensation of Si 
into a metallic (or other featureless) phase (e.g. FeSi). 
The predicted path by which MgSiO$_3$ forms is a reaction of Mg$_2$SiO$_4$ 
with SiO gas, requiring an additional oxygen atom. 
Consequently, high C/O may limit the conversion of Mg$_2$SiO$_4$ into MgSiO$_3$
and preserve a silica-poor mineralogy.
The condensation of FeSi is predicted for C/O ratios $>$0.65 
\citep[][]{fg02,lf99}. 
As C/O approaches unity, T$_{\rm dust}$ for the Mg-silicates decreases rapidly 
\citep{lf97} and Si is more apt to be incorporated into FeSi \citep{fg02}.
For C/O $>$ 0.85, the Mg-silicate dust formation 
temperature drops to below that of FeSi \citep{fg02}, 
which results in sequestration of Si and increases the Mg/Si ratio of the gas 
from which the silicates form. 
Therefore, there are at least two mechanisms by which higher
C/O ratios may be able to explain the composition of I17495.

Based on the preceding arguments we suggest that a high C/O ratio is the cause 
of the unusual mineralogy of I17495.
Our hypothesis predicts several observable correlations:
(1) between the strength of SiO absorption and C/O ratio in low $\dot{M}$ 
stars; 
(2) between SiO absorption and the relative abundances of crystalline 
forsterite and enstatite in high $\dot{M}$ stars; and
(3) between C/O and the peak positions/FWHM of the classic 10$\mu$m (emission) 
feature. 
As discussed in \S~\ref{RTmod} the peak position and FWHM of the 10$\mu$m 
feature depend on the composition of the amorphous silicate.

\section{The effect of source morphology}
\label{morph}

\citet{molster99,molster02} found a correlation between 
AGB/post-AGB star crystallinity and source morphology, with ``disk-like'' 
sources having a higher degree of crystallinity than spherical sources.
\citet{molster99} suggested that this correlation is due to low-temperature 
crystallization in a long-lived disk. 
However, the inference of a disk around highly crystalline sources comes from 
imaging and the general shape of the SED, which cannot distinguish between a 
Keplerian disk and a toroidal outflow. There is evidence that the last stages 
of mass loss from AGB stars become increasing toroidal, and that this effect 
is stronger the more massive the progenitor star 
\citep[][and references therein]{ds06}. In this case, 
the density/pressure in the dust-forming regions would be enhanced further by 
the concentration of material into the star's equatorial region, leading to 
conditions more apt to form crystalline structures. This in turn would 
correlate degree of crystallinity with initial mass of the star.
Only massive AGB stars would produce significant amounts of crystalline 
material, and these are significantly rarer than low mass AGB stars.
This resolves the conundrum of the ISM (having low crystallinity) being 
enriched by high crystallinity sources (see \S~\ref{intro}).

\section{Conclusions}

We have presented the first crystalline silicate absorption feature observed 
to date.
The spectrum of I17495 is enigmatic and suggests that its dust is dominated by 
Mg$_2$SiO$_4$ in both the crystalline and amorphous phases.
We have confirmed that high mass-loss rates should produce more crystalline 
material than low mass-loss rates; and that the crystalline mineralogy should 
be increasingly dominated by forsterite as mass-loss rates increase.
The most likely factor controlling both crystallinity and mineralogy is the 
C/O ratio.
We suggest that the correlation between crystallinity, mass-loss rate and 
initial stellar mass mitigates the problem of the very different crystal 
fractions observed in some AGB stars and in the ISM. Only the rarer, higher 
mass AGB stars contribute a significant amount of crystalline material to the 
ISM.

\acknowledgements
We are grateful to H.-P. Gail, whose comments significantly improved this 
paper. This work is supported by
NSF CAREER AST-0642991 (for A.K.S.) and
NSF CAREER EAR-0748411 (for A.G.W.).





\clearpage


\clearpage



\end{document}